\documentclass[a4paper,10pt]{article}
\usepackage{color}
\usepackage{hyperref}
\usepackage{amssymb}
\usepackage{amsmath}
\usepackage{amsfonts}
\usepackage{euscript}
\usepackage[all]{xy}

\newcommand{\Proof}{\textit{Proof:}}
\newcommand{\Sproof}{\textit{Sketch proof:}}
\newcommand{\ctory}{conceptory}
\newcommand{\ctories}{conceptories}
\newcommand{\ctorial}{conceptorial}

\newcommand{\Ctories}{Conceptories}
\newcommand{\flatt}{\textnormal{\textbf{f}}-lattice}
\newcommand{\flatts}{\textnormal{\textbf{f}}-lattices}

\newcommand{\tdcat}{\textnormal{\textbf{f}}-category}
\newcommand{\tdcats}{\textnormal{\textbf{f}}-categories}
\newcommand{\Tdcat}{\textnormal{\textbf{f}}-category}
\newcommand{\Tdcats}{\textnormal{\textbf{f}}-categories}

\newcommand{\catc}{\mathbb{C}}

\newcommand{\id}{\mathrm{id}}

\newcommand{\dom}{\mathrm{dom}}
\newcommand{\cod}{\mathrm{cod}}
\newcommand{\Ob}{\mathrm{Ob}}
\newcommand{\Mor}{\mathrm{Mor}}
\newcommand{\ifonly}{\textsl{iff}}
\newcommand{\Hom}{\mathrm{Hom}}

\title{Towards arrow-theoretic semantics of ontologies: \ctories{}}
\date{15 July 2010}
\author{Osman Bineev\\\small{\href{mailto:bineev@gmail.com}{\emph{bineev@gmail.com}}}}

\newtheorem{prop}{Proposition}[section]
\newtheorem{cor}{Corollary}[section]
\newtheorem{hyp}{Hypothesis}[section]

\begin{document}

\maketitle
\begin{abstract}
In context of efforts of composing category-theoretic and logical methods in the area of knowledge representation we propose the notion of \emph{\ctory{}}. We consider intersection/union and other constructions in \ctories{} as expressive alternative to category-theoretic (co)limits and show they have features similar to (pro-, in-)jections. Then we briefly discuss approaches to development of formal systems built on the base of conceptories and describe possible application of such system to the specific ontology.
\end{abstract}

\tableofcontents

\section{Introduction}

Ontologies \cite{handbook} are used in computer science for representing and sharing know\-ledge about the real world. Usually ontological structures are described in terms of classes (of things) and relationships (between things). This is rather similar to category-theoretic notions of objects and morphisms (see \cite{barrwells, maclane} for information about the algebraic category theory). Since the category theory already brings us many benefits in other areas of computer science, it is desirable to find arrow-theoretic approaches in the area of knowledge representation.

Some authors proposed category-theoretic techniques helpful in different aspects of knowledge representation\cite{usect1, usect2}. Usually they operate with (co)limits that are convenient for merging and interoperating between \emph{existing} models and metamodels. Our aim is to find a category-theoretic tools that would be useful for \emph{description} of ontological models from the very beginning.

In order of informal discussion, consider relations between people, like following:
\[
\xy
(-25,0)*+{Managers}="4";
(25,0)*+{Employees}="5";
{\ar@{->}^{management} "4";"5"};
\endxy
\]
Here both `\emph{Managers}' and `\textit{Employees}' are classes of individuals (within presumed classification) and `\textit{management}' can be considered as relation, something similar to morphism in category $\mathbf{Rel}$ of sets and relations. We can also consider some wider relations, for example `co-working':
\[
\xy
(-25,0)*+{Managers}="4";
(25,0)*+{Employees}="5";
{\ar@/^1.65pc/^{co-working} "4";"5"};
{\ar@/_1.65pc/_{management} "4";"5"};
{\ar@{=>} (0,-5)*{};(0,5)*{}} ;
\endxy
\]
Here the bold arrow shows something new: the `co-working' relation includes `management' (considered as simply institutional relation between people, not a process!).

But we also know how to express this fact in the language of category theory. In order to consider `arrows between arrows' we can use 2-categories \cite{maclane}.

Something really interesting happens when we consider arrow-theoretically the following picture:
\[
\xy
(-25,0)*+{Managers}="4";
(25,10)*+{Employees}="5";
(25,-10)*+{Engineers}="6";
{\ar@/^0.1pc/@{->}^{management} "4";"5"};
{\ar@/_0.1pc/@{->}_{management} "4";"6"};
{\ar@{=>} (0,-3)*{};(0,3)*{}} ;
\endxy
\]
This picture is intuitively correct: the `management of employees' includes, in some sense, `management of engineers'. But the problem with this picture is that targets of arrows connected by bold arrow should coincide, if we wish to continue using the theory of 2-categories.

Experienced mathematician would say there is the arrow between `\emph{Engineers}' and `\emph{Employees}' that makes the picture correct in 2-categories. However from the point of view of knowledge representation this is mistake: there is no significant relationship, like `\emph{management}' or `\emph{co-working}' or something else. But certainly there is some inclusion of the other nature.

Starting from this point we develop the theory of \ctories{}, where it is possible to express such situations and other things convenient in knowledge representation problems.

\Ctories{} and informal description of their language, together with discussion of possible application to the ontology of international standard ISO 15926, are introduced in the \hyperlink{ctories}{section 4} of this paper; reader who is not interested in mathematical details may skip to there.

\section{\Tdcats}

Before introducing \ctories{} we should define wider notion of \textbf{\tdcat{}}. The definition is rather complicated, but gives us almost all the instruments needed to work with ontologies. Besides, this complexity is typical for $(\geqslant\!2)$-dimensional constructions in category theory.

\subsection*{What is \tdcat?}
\Tdcat{} $\catc$ consists of

\begin{itemize}
\item Category $\catc_1$ with objects ({\bf 0-cells} of $\catc$) denoted $A, B, A_1, A_2, ...$ and arrows ({\bf 1-cells} of $\catc$) denoted $f, g, f_1, f_2, ...$ .
\[
\xy
(-10,0)*+{A}="4";
(10,0)*+{B}="5";
{\ar@{->}^{f} "4";"5"};
\endxy
\]
\item Category $\catc_v$ where objects are 1-cells of $\catc$ and arrows ({\bf 2-cells} of $\catc$) are denoted $\alpha, \beta, \alpha_1, \alpha_2, ...$ . Note that we do not require domain and codomain of 2-cell to be \emph{parallel} 1-cells. Composition of arrows in $\mathbb{C}_v$ is called \textbf{vertical} composition of 2-cells.
\[
\xy
(-20,-10)*+{A_1}="2";
(20,-10)*+{B_1}="3";
(-20,0)*+{A_2}="4";
(20,0)*+{B_2}="5";
(-20,10)*+{A_3}="6";
(20,10)*+{B_3}="7";
{\ar@{->}_{f_1} "2";"3"};
{\ar@{->}^{f_2} "4";"5"};
{\ar@{->}^{f_3} "6";"7"};
{\ar@{=>}^<<<{\scriptstyle \alpha} (-5,-9)*{};(-5,-1)*{}} ;
{\ar@{=>}^<<<{\scriptstyle \beta} (-5,1)*{};(-5,9)*{}} ;
{\ar@{=>}_<<<{\beta \circ \alpha} (5,-9)*{};(5,9)*{}} ;
\endxy
\]
\item For any pairs $(f_1, f_2)$, $(g_1, g_2)$ of composable 1-cells and for any 2-cells $\alpha{:}\,f_1 \to g_1$, $\beta{:}\,f_2 \to g_2$ the associative {\bf horizontal} composition $$\beta \star \alpha{:}\;f_2\circ f_1 \to g_2 \circ g_1,$$
\[
\xy
(-60,-5)*+{A_1}="2";
(-40,-5)*+{B_1}="3";
(-20,-5)*+{C_1}="7";
(-60,5)*+{A_2}="4";
(-40,5)*+{B_2}="5";
(-20,5)*+{C_2}="8";
(0,-5)*+{A_1}="9";
(40,-5)*+{C_1}="10";
(0,5)*+{A_2}="11";
(40,5)*+{C_2}="12";
{\ar@{->}_{f_1} "2";"3"};
{\ar@{->}^{g_1} "4";"5"};
{\ar@{->}_{f_2} "3";"7"};
{\ar@{->}^{g_2} "5";"8"};
{\ar@{->}_{f_2\circ f_1} "9";"10"};
{\ar@{->}^{g_2\circ g_1} "11";"12"};
{\ar@{=>}^<<<{\scriptstyle \alpha} (-50,-3)*{};(-50,3)*{}} ;
{\ar@{=>}^<<<{\scriptstyle \beta} (-30,-3)*{};(-30,3)*{}} ;
{\ar@{=>}^<<<{\scriptstyle \beta \star \alpha} (20,-3)*{};(20,3)*{}} ;
\endxy
\]
subject to the following \textbf{interchange law}:
\begin{equation} \label{interchange}
(\alpha_2\star \beta_2)\circ(\alpha_1\star \beta_1) = (\alpha_2\circ \alpha_1)\star(\beta_2 \circ \beta_1)
\end{equation}
\item For any 2-cells $\alpha{:}\,\id_{A'}\to\id_{A}$ and $\beta{:}\,\id_{B'}\to\id_{B}$ operations $\nabla_{\alpha, \beta}$ and $\triangle_{\alpha, \beta}$ stated as follow.

Denote $\catc_{\alpha,\beta}$ the subcategory of $\catc_v$, consisting of full subcategories\newline$\Hom_{\catc_1}(A, B)$,  $\Hom_{\catc_1}(A', B')$ and all 2-cells $\gamma$ such that
\begin{equation}\label{nabladef}
\gamma \star \alpha = \beta \star \gamma = \gamma
\end{equation} Consider inclusion functors: $$\mathbf{U}{:}\,\Hom_{\catc_1}(A, B)\to \catc_{\alpha,\beta}$$ $$\mathbf{U'}{:}\,\Hom_{\catc_1}(A', B')\to \catc_{\alpha,\beta}.$$ In category $\catc_{\alpha,\beta}$ operation $\nabla_{\alpha, \beta}$ provides for any 1-cell $f{:}\,A\to B$ universal arrow\cite{maclane} from $\mathbf U'$ to $f$.
\[
\xy
(-20,10)*+{A}="2";
(20,10)*+{B}="3";
(-20,-10)*+{A'}="4";
(20,-10)*+{B'}="5";
{\ar@{->}^{f} "2";"3"};
{\ar@{->} "4";"5"};
{\ar@{=>}^<<<{\alpha} (-20,-7)*{};(-20,7)*{}} ;
{\ar@{=>}^<<<{\beta} (20,-7)*{};(20,7)*{}} ;
{\ar@{=>}^<<<{\nabla_{\alpha, \beta}(f)} (0,-7)*{};(0,7)*{}} ;
\endxy
\]
And operation $\triangle_{\alpha, \beta}$ provides for any 1-cell $f'{:}\,A'\to B'$ universal arrow from $f'$ to $\mathbf U$.
\[
\xy
(-20,-10)*+{A'}="2";
(20,-10)*+{B'}="3";
(-20,10)*+{A}="4";
(20,10)*+{B}="5";
{\ar@{->}_{f'} "2";"3"};
{\ar@{->} "4";"5"};
{\ar@{=>}^<<<{\alpha} (-20,-7)*{};(-20,7)*{}} ;
{\ar@{=>}^<<<{\beta} (20,-7)*{};(20,7)*{}} ;
{\ar@{=>}^>>>{\triangle_{\alpha, \beta}(f')} (0,-7)*{};(0,7)*{}} ;
\endxy
\]
Operations $\nabla$ and $\triangle$ are subjects to following restrictions:
\begin{align}
\label{ndistDown} &\nabla_{\id_{\id},\beta}(f) \star \nabla_{\alpha, \id_{\id}}(g) = \nabla_{\alpha, \beta}(f\circ g)\\
\label{ndistUp} &\triangle_{\id_{\id},\beta}(f) \star \triangle_{\alpha, \id_{\id}}(g) = \triangle_{\alpha, \beta}(f\circ g)
\end{align}
\end{itemize}

That's all about the definition of \tdcat. In addition, denote
\begin{align*}
&f{\downarrow_{\alpha, \beta}} \equiv \dom_{\scriptscriptstyle \catc_v}(\nabla_{\alpha, \beta}(f))\quad
&f{\uparrow^{\alpha, \beta}} \equiv \cod_{\scriptscriptstyle \catc_v}(\triangle_{\alpha, \beta}(f))\\
&f{\downharpoonleft_{\alpha}} \equiv \dom_{\scriptscriptstyle \catc_v}(\nabla_{\alpha, \id_\id}(f))\quad
&f{\upharpoonleft^{\alpha}} \equiv \cod_{\scriptscriptstyle \catc_v}(\triangle_{\alpha, \id_\id}(f))\\
&f{\downharpoonright_{\beta}} \equiv \dom_{\scriptscriptstyle \catc_v}(\nabla_{\id_\id, \beta}(f))\quad
&f{\upharpoonright^{\beta}} \equiv \cod_{\scriptscriptstyle \catc_v}(\triangle_{\id_\id, \beta}(f))
\end{align*}

\newcommand{\frel}{\mathbf{^f{Rel}}}

\subsection*{Example: $\frel$}

Let $X$ be a set and $\mathbf{Rel}(X)$ be a full subcategory of $\mathbf{Rel}$ generated by all subsets of $X$. $\mathbf{Rel}(X)$ can be turned to \tdcat{} $\frel(X)$ taking $\frel(X)_1 = \mathbf{Rel}(X)$.

Recall that morphism $f{:}\,A\to B$ of $\mathbf{Rel}$ is a triple $\langle{r_f}, A, B\rangle$ where $A$ and $B$ are sets and $r_f\subseteq A\times B$. The only 2-cell $f\to g$ exists \ifonly{} $r_f\subseteq r_g$. Operations $\uparrow$ and $\downarrow$ for $f$ are introduces as follows:
\begin{align*}
&\langle{r_f}, A, B\rangle{\downarrow_{A',B'}} \equiv \langle{(A'\times B')\cap r_f}, A', B'\rangle,\,\mathrm{where}\,A'\subseteq A, B'\subseteq B\\
&\langle{r_f}, A, B\rangle{\uparrow^{A',B'}} \equiv  \langle{r_f}, A', B'\rangle,\,\mathrm{where}\,A\subseteq A', B\subseteq B'
\end{align*}
Operation `$\star$' takes place: if $r_{f_1}\subseteq r_{g_1}$ and $r_{f_2}\subseteq r_{g_2}$ then obviously $r_{f_2}\circ r_{f_1} \subseteq r_{g_2}\circ r_{g_1}$. It is associative and satisfies interchange law (\ref{interchange}) just because there is no alternative.

\subsection*{Functoriality of $\downarrow$ and $\uparrow$}
Note that for given 0-cell $A$ the category $\Hom_{\scriptscriptstyle \catc_1}(A, A)$ can be considered as monoidal, where action of tensor on objects is given by $\circ$ (composition of 1-cells), on arrows by $\star$ (horizontal composition of 2-cells) and tensor unit $\id_A$. Denote this monoidal category $\mathbf{H}_A$.

\begin{prop}\label{semidist}
For any 2-cells $\alpha{:}\,\id_{A'}\to \id_A$, $\beta{:}\,\id_{B'}\to \id_B$ and $\gamma{:}\,\id_{C'}\to \id_C$, any 1-cells $f{:}\,A\to B$, $g{:}\,B\to C$, $f'{:}\,A'\to B'$, $g'{:}\,B'\to C'$
\[
\xy
(-30,-10)*+{A'}="1";
(-30,10)*+{A}="2";
(-10,-10)*+{B'}="3";
(-10,10)*+{B}="4";
(10,-10)*+{C'}="5";
(10,10)*+{C}="6";
{\ar@{=>}_{\alpha} "1";"2"};
{\ar@{=>}_{\beta} "3";"4"};
{\ar@{=>}_{\gamma} "5";"6"};
{\ar@{->}_{f'} "1";"3"};
{\ar@{->}_{g'} "3";"5"};
{\ar@{->}_{f} "2";"4"};
{\ar@{->}_{g} "4";"6"};
\endxy
\]
there is at least one 2-cell in $\Hom_{\scriptscriptstyle \catc_v}(g{\downarrow_{\beta, \gamma}}\circ f{\downarrow_{\alpha, \beta}}, (g\circ f){\downarrow_{\alpha, \gamma}})$ and at least one 2-cell in $\Hom_{\scriptscriptstyle \catc_v}((g'\circ f'){\uparrow^{\alpha, \gamma}}, g'{\uparrow^{\beta, \gamma}}\circ f'{\uparrow^{\alpha, \beta}})$.
\end{prop}
\Proof{} First 2-cell exists by universality of $\nabla_{\alpha, \gamma}$ as a factorizing arrow for $\nabla_{\beta, \gamma}(g)\star\nabla_{\alpha, \beta}(f)$. And second by universality of $\triangle_{\alpha, \gamma}$ as a factorizing arrow for $\triangle_{\beta, \gamma}(g')\star\triangle_{\alpha, \beta}(f')$.$\square$

\begin{prop}\label{functoriality}\textnormal{\textbf{(Functoriality of $\downarrow$ and $\uparrow$)}}
\begin{enumerate}
\item For given 2-cells $\alpha{:}\,\id_{A'}\to \id_A$ and $\beta{:}\,\id_{B'}\to \id_B$ operation $\downarrow_{\alpha,\beta}$ induces functor $$\mathbf{\downarrow}_{\alpha,\beta}{:}\,\Hom_{\scriptscriptstyle \catc_1}(A, B) \to \Hom_{\scriptscriptstyle \catc_1}(A', B').$$
\item For given 2-cell $\alpha{:}\,\id_{A'}\to \id_A$ the functor $\mathbf{\downarrow}_{\alpha,\alpha}$ has monoidal structure:
$$(\mathbf{\downarrow}_{\alpha,\alpha}, \phi_\alpha, \phi_{\alpha,0}){:}\,\mathbf{H}_A \to \mathbf{H}_{A'}.$$
\item Same for $\uparrow$, except that $\uparrow^{\alpha, \alpha}$ is comonoidal functor.
\end{enumerate}
\end{prop}
Proof can be found in the \hyperlink{appA}{\textbf{Appendix A}} of this paper.

It is obvious that restricting the class of 2-cells to those between \emph{parallel} 1-cells in \tdcat{} gives us some 2-category. Therefore we may adopt some notions of 2-categories like, for example, \textbf{adjunctions}\cite{maclane}.

Although research in this direction is not in goals of this paper, we would like to state hypothesis, that could unite both propositions \ref{semidist} and \ref{functoriality} in one:

\begin{hyp}\textnormal{\textbf{(2-functoriality of $\downarrow$ and $\uparrow$)}}\\
For any function $\xi{:}\,\Ob(\catc_1)\to\Mor(\catc_v)$, such that always $$\dom(\xi(A))=\id_A \land \exists{B}:\cod(\xi(A))=\id_B$$ the operation $f\mapsto f{\downarrow_{\xi(\dom(f)), \xi(\cod(f))}}$ has structure of 2-endofunctor over $\catc$ taken as 2-category.
\end{hyp}

\section{\flatts{}}
Recall the notion of \textbf{thin} category: it is a category having at most one morphism in each homset. The \tdcat{} $\catc$ is called \textbf{\flatt} \ifonly{} $\catc_v$ is thin category with finite products and finite coproducts, where \emph{bifunctors} of binary coproduct `$\cup$' and of binary product `$\cap$' have following properties:
\begin{align}
&\id_{\dom(f\cup g)} = \id_{\dom(f)}\cup \id_{\dom(g)}\\
&\id_{\cod(f\cup g)} = \id_{\cod(f)}\cup \id_{\cod(g)}\\
&\id_{\dom(f\cap g)} = \id_{\dom(f)}\cap \id_{\dom(g)}\\
&\id_{\cod(f\cap g)} = \id_{\cod(f)}\cap \id_{\cod(g)}
\end{align}
We will write $A\cup B$ instead of $\dom(\id_A \cup \id_B)$ and $A\cap B$ instead of $\dom(\id_A \cap \id_B)$. It is also convenient to use expression $A \leqslant B$ (for 0-cells $A$ and $B$) as shortened form of $\id_A \leqslant \id_B$. 

Since, when work with \flatts{}, there is at most one 2-cell in each $\catc_v$-homset, we will use domains (codomains) of id's of 2-cells in indices of operation $\nabla$ ($\triangle$). For instance, $\nabla_{A,B}(f)$ is defined when $A \leqslant \dom(f)$ and $B \leqslant \cod(f)$. And same for $\downarrow, \uparrow, \downharpoonleft, \upharpoonleft, \downharpoonright, \upharpoonright$.

\subsection*{Operations $\downharpoonleft$ and $\upharpoonright$}

Call the predicate $X$ over 1-cells \textbf{$\downharpoonleft$-preserved} (\textbf{$\upharpoonright$-preserved}) \ifonly{} for any $f{:}\,A \to B$ proposition $X(f{\downharpoonleft_{A'}})$ (proposition $X(f{\upharpoonright^{B'}})$) follows from $X(f)$ whenever $A'\leqslant A$ (whenever $B \leqslant B'$). Using our results \ref{semidist} and \ref{functoriality} it's easy to prove the following proposition:

\begin{prop} \label{updownprop} In any \flatt{} following predicates (of argument $f{:}\,A\to B$) are both $\downharpoonleft$-preserved and $\upharpoonright$-preserved:
\begin{enumerate}
\item $\exists{g}: \id_A\leqslant f\circ g$
\item $\exists{g}: g\circ f \leqslant \id_B$
\item $\exists{g}: (\id_A\leqslant f\circ g) \land (f\circ g\circ f = f)$
\item $\exists{g}: (g\circ f \leqslant \id_B) \land (g\circ f\circ g = g)$ 
\end{enumerate}
\end{prop}
\Sproof{} $\id_{A'} \leqslant (\id_A){\downarrow_{A', A'}}$ follows from existing of $\phi_{A',0}$ in monoidal structure of functor $\downarrow_{A', A'}$ and $(\id_B){\uparrow^{B', B'}} \leqslant \id_{B'}$ dually from comonoidal structure. Next, $g{\downharpoonright_{A'}}\circ f{\downharpoonleft_{A'}} = (g \circ f){\downarrow_{A',A'}}$ by property (\ref{ndistDown}) and $f{\upharpoonright_{B'}}\circ g{\upharpoonleft_{B'}} = (f \circ g){\uparrow_{B',B'}}$ by property  (\ref{ndistUp}) from the definition of \tdcat{}. These facts together with functoriality of $\uparrow$ and $\downarrow$ are enough to prove $\downharpoonleft$-preserving of the first inequality and $\upharpoonright$-preserving of the second.

Then for $\upharpoonright$-preserving of $\id_A\leqslant f\circ g$ and $\downharpoonleft$-preserving of $g\circ f \leqslant \id_B$ we need the proposition \ref{semidist} in order to prove $g{\downharpoonright_{A'}}\circ f{\downharpoonleft_{A'}} \leqslant (g \circ f){\downarrow_{A',A'}}$ and $f{\upharpoonright_{B'}}\circ g{\upharpoonleft_{B'}} \geqslant (f \circ g){\uparrow_{B',B'}}$.

Preserving of the last two inequalities follows easily from the properties (\ref{ndistDown}) and (\ref{ndistUp}) of \tdcats{}, from proposition \ref{semidist} and using the  $\star$ operation for 2-cells.$\square$

Call the 1-cell $f{:}\,A\to B$ a \textbf{map} \ifonly{} it has right adjoint. As easy to see, in \flatts{} it is equal to existing of $g{:}\,B\to A$ such that
$$(\id_A\leqslant f\circ g)\land (g\circ f \leqslant \id_B)\land (f\circ g\circ f = f)\land (g\circ f\circ g = g)$$

\begin{cor}\label{mapstab}The predicate `to be a map' is both $\downharpoonleft$-preserved and $\upharpoonright$-preserved.\end{cor}
\begin{cor}\label{idupdown}For any 0-cells $A \leqslant B$ both $(\id_{A}){\upharpoonright^{B}}$ and $(\id_{B}){\downharpoonleft_{A}}$ are maps.\end{cor}

Two last results give us reasons of importance of operations $\downharpoonleft$ and $\upharpoonright$: they both preserve maps and make maps from $\id$'s. Another reason is that these operations are enough to introduce \textit{hybrid composition} `$\ast$' of 2-cells and 1-cells (again preserving maps!), not only in \flatts{} but in general \tdcats{}, as shown on the following picture:
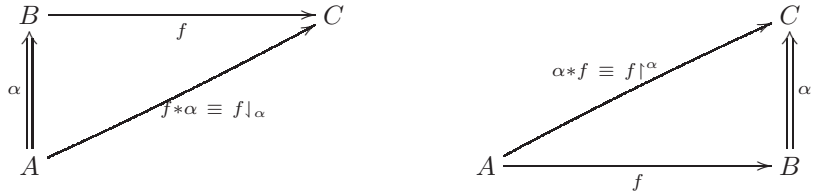
\[
\xy
(-60,-10)*+{A}="1";
(-60,10)*+{B}="2";
(-20,10)*+{C}="3";
(0,-10)*+{A}="4";
(40,-10)*+{B}="5";
(40,10)*+{C}="6";
{\ar@{->}_{f} "2";"3"};
{\ar@/_0.1pc/@{->}_{f\ast\alpha\ \equiv\ f{\downharpoonleft_{\alpha}}} "1";"3"};
{\ar@{=>}^{\alpha} "1";"2"};
{\ar@{->}_{f} "4";"5"};
{\ar@/^0.1pc/@{->}^{\alpha\ast f\ \equiv\ f{\upharpoonright^{\alpha}}} "4";"6"};
{\ar@{=>}_{\alpha} "5";"6"};
\endxy
\]

\hypertarget{invol}{}
\subsection*{Involution $(-)^\circ$}

In some cases it is useful to have for each 1-cell $f{:}\,A\to B$ it's `transposition' $f^\circ{:}\,B\to A$. Maybe definition of \flatt{} should be extended by such operation, by analogy with \emph{allegories}\cite{elephant}. In order to do that we should equip our \flatt{} $\catc$ with involutive contravariant endofunctor $(-)^\circ$ over $\catc_1$ such that $\dom(f^\circ) = \cod(f)$ and $\cod(f^\circ) = \dom(f)$. In addition it should satisfy the \emph{modular law}:
$$(f\circ g)\cap h \leqslant (f\cap (h\circ g^\circ))\circ g$$
The useful property of such involution is that whenever $f$ has right adjoint $g$ it coincides with $f^\circ$: $g = f^\circ$ \cite{elephant}.

\subsection*{Constructions in \flatts}

Following constructions can be useful in real tasks of knowledge representation:
\begin{itemize}
\item $A\cap B$ and $A \cup B$:
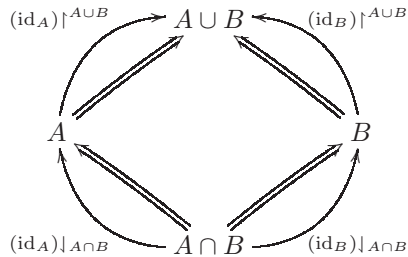
\[
\xy
(-20,0)*+{A}="2";
(20,0)*+{B}="3";
(0,15)*+{A\cup B}="4";
(0,-15)*+{A\cap B}="5";
{\ar@/^0.1pc/@{=>}^{} "2";"4"};
{\ar@/_0.1pc/@{=>}_{} "3";"4"};
{\ar@/^0.1pc/@{<=}^{} "2";"5"};
{\ar@/_0.1pc/@{<=}_{} "3";"5"};
{\ar@/^1.5pc/@{->}^{(\id_A){\upharpoonright^{A\cup B}}} "2";"4"};
{\ar@/_1.5pc/@{->}_{(\id_B){\upharpoonright^{A\cup B}}} "3";"4"};
{\ar@/_1.5pc/@{<-}_{(\id_A){\downharpoonleft_{A\cap B}}} "2";"5"};
{\ar@/^1.5pc/@{<-}^{(\id_B){\downharpoonleft_{A\cap B}}} "3";"5"};
\endxy
\]
Note that, by analogy with categorical binary products and sums (coproducts) \cite{barrwells}, we have `projections'
\begin{align*}
pr_A \equiv (\id_A){\downharpoonleft_{A\cap B}}\\
pr_B \equiv (\id_B){\downharpoonleft_{A\cap B}}
\end{align*}
and `injections'
\begin{align*}
in_A \equiv (\id_A){\upharpoonright^{A\cup B}}\\
in_B \equiv (\id_B){\upharpoonright^{A\cup B}}
\end{align*}
By corollary \ref{idupdown} all of them are maps. Surely we may consider other operations ($\downharpoonright$ and $\upharpoonleft$) but they are not bounded to produce maps.
\item Intuitively, if ordering in \flatt{} is understood as subclassing, we can understand $\downharpoonleft$ as \emph{inheritance} of features, relationships, methods, etc. For example, when $A$ and $B$ have 1-cells $f{:}\,A\to A'$ and $f{:}\,B\to B'$, $A\cap B$ \emph{inherits} both via $\downharpoonleft$:
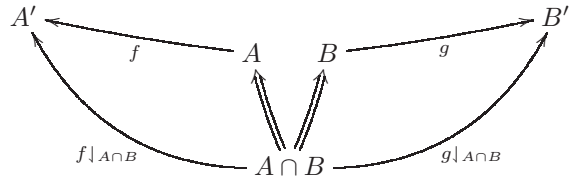
\[
\xy
(-5,0)*+{A}="2";
(-35,5)*+{A'}="3";
(5,0)*+{B}="4";
(35,5)*+{B'}="5";
(0,-15)*+{A\cap B}="6";
{\ar@/^0.1pc/@{->}^{f} "2";"3"};
{\ar@/_0.1pc/@{->}_{g} "4";"5"};
{\ar@/^0.1pc/@{=>}^{} "6";"2"};
{\ar@/_0.1pc/@{=>}_{} "6";"4"};
{\ar@/^1.7pc/@{->}^{f{\downharpoonleft_{A\cap B}}} "6";"3"};
{\ar@/_1.7pc/@{->}_{g{\downharpoonleft_{A\cap B}}} "6";"5"};
\endxy
\]
Again, if $f$ and $g$ are maps, or 1-cells with some properties, mentioned in \ref{updownprop}, these properties will be preserved.
\item Consider situation when $A$ and $B$ have 1-cells $f{:}\,A\to C$ and $g{:}\,B\to C$ and we wish to inherit them and unite in one 1-cell. We could use $f\cap g$, since by definition $\dom(f\cap g) = A\cap B$. But this is not convenient, because $f\cap g$ is not guaranteed to be a map. In fact we need another operation $f \cap_{\scriptscriptstyle C} g$ such that $$(f{\downharpoonleft_{\dom(f \cap_{\scriptscriptstyle C} g)}} = f \cap_{\scriptscriptstyle C} g) \land (g{\downharpoonleft_{\dom(f \cap_{\scriptscriptstyle C} g)}} = f \cap_{\scriptscriptstyle C} g)$$ and $h \leqslant f \cap_{\scriptscriptstyle C} g$ for any other $h$ with these properties. 
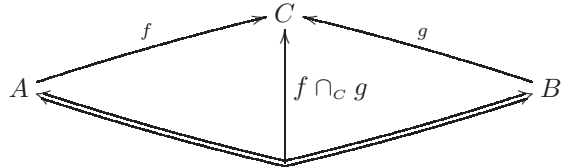
\[
\xy
(-35,0)*+{A}="2";
(35,0)*+{B}="3";
(0,-10)*+{}="4";
(0,10)*+{C}="5";
{\ar@/^0.1pc/@{->}^{f} "2";"5"};
{\ar@/_0.1pc/@{->}_{g} "3";"5"};
{\ar@/^0.1pc/@{=>}^{} "4";"2"};
{\ar@/_0.1pc/@{=>}_{} "4";"3"};
{\ar@{->}_{\textstyle {f \cap_{\scriptscriptstyle C} g}} "4";"5"};
\endxy
\]
Call it \textbf{logical pullback}. Logical pullbacks preserve good properties, but as far as we can see this construction does not follow from the definition of \flatt{}. This operation is similar to pullback square \cite{barrwells} from the usual category theory, so it is good subject for the future research.
\end{itemize}

\hypertarget{ctories}{}
\section{\Ctories{} and their language}

If a \flatt{} is, in addition, complete heyting (boolean) algebra, then we call it \textbf{\ctory{}} (\textbf{boolean \ctory{}}). But this section is dedicated not so much to mathematical properties of \ctories{}, but to (rather \emph{informal}) description of possible formal system that this notion induces and that could be used in ontological applications.

\subsection*{Formal system}
Thus, 0-cells ($A, B, C, A_1, A_2, ...$) of \ctory{} become \emph{classes} ($\mathtt{A, B, C, A_1, A_2, ...}$) of our ontological language and 1-cells ($f, g, h, f_1, f_2, ...$) become typed \emph{relationships} ($\mathtt{f, g, h, f_1, f_2, ...}$). As before we may describe domain and codomain of relationship, for example $\mathtt{f{:}\,A\to B}$, and compose them, $\mathtt{f\circ g}$, in associatie way. The apparatus of heyting or boolean algebras gives us the full set of logical connectives over relationships and classes, with the usual collection of axioms. It is not necessary to describe them here --- we are going to concentrate on specific axioms and rules of \ctories{}.

First introduce some auxiliary axioms:
\begin{align*}
\mathtt{\frac{\underline{\id_A\Rightarrow \id_B}}{A\Rightarrow B}(\id{\Rightarrow})},
\end{align*}
where `$\Rightarrow$' denotes logical implication;
\begin{align*}
\mathtt{\frac{\underline{f{:}\,A\to B\quad A'\Rightarrow A\quad B'\Rightarrow B}}{def_{A,B}(f{\downarrow_{A',B'}})}(def{\downarrow})}\quad
\mathtt{\frac{\underline{f{:}\,A\to B\quad A\Rightarrow A'\quad B\Rightarrow B'}}{def_{A,B}(f{\uparrow^{A',B'}})}(def{\uparrow})}
\end{align*}
Now describe specific axioms and rules of \tdcats{}. The first one comes from the `$\star$' operation over 2-cells:
$$\mathtt{\frac{f_1 \Rightarrow g_1\qquad f_2 \Rightarrow g_2}{f_2\circ f_1\Rightarrow g_2\circ g_1} }(\star)$$
Note that associativity of `$\star$' and interchange law (\ref{interchange}) are guaranteed, since there are no alternatives.

Then we have several axioms for $\mathtt{\downarrow}$ and $\mathtt{\uparrow}$, as consequences of definitions of $\nabla$ and $\triangle$:
\begin{align*}
&\mathtt{\frac{def_{A,B}(f{\downarrow_{A',B'}})}{f{\downarrow_{A',B'} \Rightarrow f}}(\downarrow)}\qquad
\mathtt{\frac{def_{A,B}(f{\downarrow_{A',B'}})\quad g'{:}\,A'\to B'\quad g'\Rightarrow f}{g'\Rightarrow f{\downarrow_{A',B'}}}(univ{\downarrow})}\\
\\
&\mathtt{\frac{def_{A,B}(f{\uparrow^{A',B'}})}{f \Rightarrow f{\uparrow^{A',B'}}}(\uparrow)}\qquad
\mathtt{\frac{def_{A,B}(f{\uparrow^{A',B'}})\quad g'{:}\,A'\to B'\quad f\Rightarrow g'}{f{\uparrow^{A',B'}}\Rightarrow g'}(univ{\uparrow})}\\
\\
&\mathtt{\frac{}{f{\downharpoonright_{B}}\circ g{\downharpoonleft_{A}} \Leftrightarrow (f\circ g){\downarrow_{A, B}}}(distrib{\downarrow})}\quad
\mathtt{\frac{}{f{\upharpoonright^{B}}\circ g{\upharpoonleft^{A}} \Leftrightarrow (f\circ g){\uparrow^{A, B}}}(distrib{\uparrow})}
\end{align*}
Sometimes we will omit $\mathtt{def_{A,B}(f{\downarrow_{A',B'}})}$ and $\mathtt{def_{A,B}(f{\uparrow_{A',B'}})}$ for simplicity of formulae. Following axioms come from the definition of \flatt{}:
\begin{align*}
&\mathtt{\frac{f{:}\,A\to B\quad g{:}\,C\to D}{(f\land g){:}\,(A\land C)\to (B\land D)}(bounds{\land})}\quad
\mathtt{\frac{f{:}\,A\to B\quad g{:}\,C\to D}{(f\lor g){:}\,(A\lor C)\to (B\lor D)}(bounds{\lor})}
\end{align*}

\newcommand{\termt}{\top}
In order to introduce \emph{elements} to our language we could add distinguished class $\mathtt{\termt}$ and use $\mathtt{x{:}A}$ and $\mathtt{f(x)}$ instead of $\mathtt{x{:}\,\termt\to A}$ and $\mathtt{f\circ x}$ correspondently. But usually we don't need such extension of language thanks to the power of algebraic representation.

Proposition \ref{semidist} and \ref{functoriality} give us following theorems:
\begin{align}
&\mathtt{g{\downarrow_{B', C'}}\circ f{\downarrow_{A',B'}} \Rightarrow (g\circ f){\downarrow_{A',C'}}}\\
&\mathtt{(g\circ f){\uparrow^{A',C'}} \Rightarrow g{\uparrow^{B', C'}}\circ f{\uparrow^{A',B'}}}\\
\nonumber\\
&\mathtt{\frac{f\Rightarrow g}{f{\downarrow_{A',B'}}\Rightarrow g{\downarrow_{A',B'}}}}\\
\nonumber\\
&\mathtt{\frac{f\Rightarrow g}{f{\uparrow^{A',B'}}\Rightarrow g{\uparrow^{A',B'}}}}\\
\nonumber\\
&\mathtt{\id_{A'}\Rightarrow (\id_A){\downarrow_{A',A'}}}\\
&\mathtt{(\id_{A'}){\uparrow^{A,A}} \Rightarrow \id_A}
\end{align}

By convention we will use sometimes $\mathtt{A{.}f{:}\,B}$ instead of $\mathtt{f{:}\,A\to B}$ and $\mathtt{A{.}f}$ instead of $\mathtt{f}$ when domain of $\mathtt{f}$ is proven to be $\mathtt{A}$.

\subsection*{Towards application to ISO 15926}

ISO 15926 \cite{iso15926} is international standard for industrial automation systems and integration. Part 2 of this standard contains description of ontology consisting of 201 entity types, with EXPRESS language. Part ISO 15926-7 describes the same information in the language of first order logic (FOL).

Texts of \href{http://www.steptools.com/sc4/archive/*checkout*/oil-and-gas/15926-0002-lifecycle_integration.exp?rev=HEAD&content-type=text/plain}{EXPRESS representation} and \href{https://www.posccaesar.org/wiki/ISO15926inFOL}{FOL representation} of the standard are available online. 

Although currently we don't have any consistency checking algorithms for \ctories{}, we hope to obtain such algorithms and below we make preliminary notes of how to describe ISO 15926 entities in \ctorial{} language.

Let $\mathtt{ab\_cd}$ is EXPRESS entity type. We present this in \ctorial{} language as class (something corresponding to 0-cell) $\mathtt{AbCd}$. For example, entity type $\mathtt{class\_of\_relationship}$ becomes \ctorial{} class $\mathtt{ClassOfRelationship}$.

Then, attributes will be presented in our language as 1-cells. For example, entity type $\mathtt{classification}$ declares two attributes: $\mathtt{classified}$ of type $\mathtt{thing}$ and $\mathtt{classifier}$ of type $\mathtt{class}$. In our language this facts are presented as follow:
\begin{align*}
\mathtt{classified{:}\,Classification\to Thing}\\
\mathtt{classifier{:}\,Classification\to Class}
\end{align*}
or, using syntactic sugar described in previous paragraph, simply
\begin{align*}
\mathtt{Classification.classified{:}\,Thing}\\
\mathtt{Classification.classifier{:}\,Class}
\end{align*}
Several types of information are represented using almost same principles as ISO 15926-7 uses in case of FOL. Subtyping among entity types is represented using implication; $\mathtt{A\Rightarrow B}$ means that $\mathtt{A}$ is subclass of $\mathtt{B}$. If entity type $\mathtt{a}$ is ABSTRACT and has immediate subtypes $\mathtt{b, c, d}$, we present this in form $$\mathtt{A\Rightarrow B\lor C\lor D}$$ The EXPRESS $\mathtt{ONEOF(a, b, c)}$ is represented with formulae
\begin{align*}
&\mathtt{\lnot(A\land(B\lor C))}\\
&\mathtt{\lnot(B\land C)}
\end{align*}
Next, by convention, if we have $\mathtt{A.f{:}\,B}$ and $\mathtt{C\Rightarrow A}$ then we understand $\mathtt{C.f}$ as $\mathtt{f{\downharpoonleft_{C}}}$. But if $\mathtt{C}$ redeclares codomain of $\mathtt{f}$ (to $\mathtt{B'}$ for instance) it should be described distinctly: $$\mathtt{C.f\Leftrightarrow(A.f){\downarrow_{\bullet,B'}}},$$ where $\bullet$ signs adopting name of domain $\mathtt{C}$ from the name of 1-cell. Here is an example from the standard:
\begin{align*}
\mathtt{UpperBoundOfNumberRange.classified}&\mathtt{\quad\Leftrightarrow}\\\mathtt{Classification.classified}&\mathtt{{\downarrow_{\bullet, ArithmeticNumber}}}
\end{align*}

Till now we didn't use any variables over elements of conceptorial classes. Cardinality constraints can also be expressed in this style using involutive operation $(-)^\circ$ (\hyperlink{invol}{described} in previous section) and in terms of proposition \ref{updownprop}. For example, cardinality constraint $[1,\ast)$ for 1-cell $\mathtt{A.f{:}\,B}$ is presented in form $$\mathtt{\id_A\Rightarrow A.f^{\circ}\circ A.f},$$ Next, the cardinality constraint $[0,1]$ is presented as $$\mathtt{A.f\circ A.f^{\circ} \Rightarrow \id_B}$$ And finally, EXPRESS UNIQUE restriction is presented as $$\mathtt{A.f^{\circ}\circ A.f\Rightarrow \id_A}$$

\hypertarget{appA}{}
\section{Appendix A}
Proof of the proposition \ref{functoriality} (functoriality of $\downarrow$ and $\uparrow$):
\begin{enumerate}
\item Consider 1-cells $f,g{:}\,A\to B$ and 2-cell $\gamma{:}\,f\to g$. With $\nabla$ we obtain 2-cells $\nabla_{\alpha, \beta}(g){:}\,g{\downarrow_{\alpha, \beta}}\to g$ and $\gamma\circ\nabla_{\alpha, \beta}(f){:}\,f{\downarrow_{\alpha, \beta}}\to g$. By universality of $\nabla$ there is unique 2-cell $\gamma'{:}\,f{\downarrow_{\alpha, \beta}}\to g{\downarrow_{\alpha, \beta}}$ such that
\begin{equation} \label{functordef}
\nabla_{\alpha, \beta}(g)\circ\gamma' = \gamma\circ\nabla_{\alpha, \beta}(f).
\end{equation}
\[
\xy
(-10,-10)*+{f{\downarrow_{\alpha, \beta}}}="2";
(10,-10)*+{g{\downarrow_{\alpha, \beta}}}="3";
(-10,10)*+{f}="4";
(10,10)*+{g}="5";
{\ar@{=>}_{\gamma'} "2";"3"};
{\ar@/_0.1pc/@{=>}_{} "2";"5"};
{\ar@{=>}^{\gamma} "4";"5"};
{\ar@{<=}_{\nabla_{\alpha, \beta}(f)} "4";"2"};
{\ar@{<=}^{\nabla_{\alpha, \beta}(g)} "5";"3"};
\endxy
\]
Define $\gamma{\downarrow_{\alpha, \beta}} \equiv \gamma'.$ It's obvious from definition that $(\id_f){\downarrow_{\alpha, \beta}} = \id_{(f{\downarrow_{\alpha, \beta}})}$. The rule $$(\gamma_1\circ\gamma_2){\downarrow_{\alpha, \beta}} = \gamma_1{\downarrow_{\alpha, \beta}}\circ\gamma_2{\downarrow_{\alpha, \beta}}$$ can be derived from commutativity of the outer square of the following diagram:
\[
\xy
(-20,-10)*+{f{\downarrow_{\alpha, \beta}}}="2";
(0,-10)*+{g{\downarrow_{\alpha, \beta}}}="3";
(20,-10)*+{h{\downarrow_{\alpha, \beta}}}="6";
(-20,10)*+{f}="4";
(0,10)*+{g}="5";
(20,10)*+{h}="7";
{\ar@{=>}_{\gamma_2'} "2";"3"};
{\ar@{=>}^{\gamma_2} "4";"5"};
{\ar@{=>}_{\gamma_1'} "3";"6"};
{\ar@{=>}^{\gamma_1} "5";"7"};
{\ar@{<=}_{\nabla_{\alpha, \beta}(f)} "4";"2"};
{\ar@{<=}^{\nabla_{\alpha, \beta}(g)} "5";"3"};
{\ar@{<=}^{\nabla_{\alpha, \beta}(h)} "7";"6"};
\endxy
\]
\item First derive $\phi_{\alpha,0}$. Both 2-cells $\alpha$ and $\nabla_{\alpha, \alpha}(\id_A)$ have codomain $\id_A$, so by universality there is unique $\gamma''{:}\,\id_A'\to(\id_A){\downarrow_{\alpha, \alpha}}$ such that 
\begin{equation} \label{mf0def}
\alpha = \nabla_{\alpha, \alpha}(\id_A)\circ\gamma''
\end{equation}
Now define $\phi_{\alpha,0} \equiv \gamma''$.
\[
\xy
(-20,-10)*+{\id_{A'}}="2";
(20,-10)*+{(\id_A){\downarrow_{\alpha, \alpha}}}="3";
(0,10)*+{\id_A}="4";
{\ar@/^0.1pc/@{=>}^{\alpha} "2";"4"};
{\ar@/_0.1pc/@{=>}_{\nabla_{\alpha, \alpha}(\id_A)} "3";"4"};
{\ar@{=>}_{\gamma''} "2";"3"};
\endxy
\]
We also need $\phi_\alpha$. Consider 1-cells $f{:}\,A\to A$ and $g{:}\,A\to A$. Together with $\nabla_{\alpha, \alpha}(f\circ g)$ we have $\nabla_{\alpha, \alpha}(f)\star\nabla_{\alpha, \alpha}(g)$ and, by universality again, the 2-cell $\gamma'''{:}\,f{\downarrow_{\alpha, \alpha}}\circ g{\downarrow_{\alpha, \alpha}}\to(f\circ g){\downarrow_{\alpha, \alpha}}$ with property
\begin{equation}\label{mfdef}
\nabla_{\alpha, \alpha}(f\circ g)\circ\gamma''' = \nabla_{\alpha, \alpha}(f)\star\nabla_{\alpha, \alpha}(g)
\end{equation}
\[
\xy
(-20,-10)*+{f{\downarrow_{\alpha, \alpha}}\circ g{\downarrow_{\alpha, \alpha}}}="2";
(20,-10)*+{(f\circ g){\downarrow_{\alpha, \alpha}}}="3";
(0,10)*+{f\circ g}="4";
{\ar@/^0.1pc/@{=>}^{\nabla_{\alpha, \alpha}(f)\star\nabla_{\alpha, \alpha}(g)} "2";"4"};
{\ar@/_0.1pc/@{=>}_{\nabla_{\alpha, \alpha}(f\circ g)} "3";"4"};
{\ar@{=>}_{\gamma'''} "2";"3"};
\endxy
\]
Define $\phi_\alpha(g, f) \equiv \gamma'''$. Naturality of $\phi_\alpha$ by $g$ means that 
\begin{equation}\label{naturality}
\phi_\alpha(g, f)\circ(\id_f{\downarrow_{\alpha, \alpha}}\star\beta'{\downarrow_{\alpha, \alpha}}) = (\id_f\star\beta'){\downarrow_{\alpha, \alpha}}\circ \phi_\alpha(h, f)
\end{equation}
for any $\beta'{:}\,h\to g$. Let's prove it. First, by (\ref{mfdef}) we have
\begin{align*}
\nabla_{\alpha, \alpha}(f\circ g)\circ\phi_\alpha(g, f)\circ(\id_f{\downarrow_{\alpha, \alpha}}\star\beta'{\downarrow_{\alpha, \alpha}}) =\\(\nabla_{\alpha, \alpha}(f)\star\nabla_{\alpha, \alpha}(g))\circ(\id_f{\downarrow_{\alpha, \alpha}}\star\beta'{\downarrow_{\alpha, \alpha}})
\end{align*}
Let's modify the right side of this equality, using (\ref{functordef}): 
$$\nabla_{\alpha, \alpha}(f\circ g)\circ\phi_\alpha(g, f)\circ(\id_f{\downarrow_{\alpha, \alpha}}\star\beta'{\downarrow_{\alpha, \alpha}}) = (\id_f\star\beta')\circ(\nabla_{\alpha, \alpha}(f)\star\nabla_{\alpha, \alpha}(h))$$
Now, again by property (\ref{mfdef}) (but with different components) $$\nabla_{\alpha, \alpha}(f\circ g)\circ\phi_\alpha(g, f)\circ(\id_f{\downarrow_{\alpha, \alpha}}\star\beta'{\downarrow_{\alpha, \alpha}}) = (\id_f\star\beta')\circ\nabla_{\alpha, \alpha}(f\circ h)\circ\phi_\alpha(h, f)$$
And by (\ref{functordef})
\begin{align*}
\nabla_{\alpha, \alpha}(f\circ g)\circ\phi_\alpha((g, f)\circ(\id_f{\downarrow_{\alpha, \alpha}}\star\beta'{\downarrow_{\alpha, \alpha}}) =\\\nabla_{\alpha, \alpha}(f\circ g)\circ(\id_f\star\beta'){\downarrow_{\alpha, \alpha}}\circ \phi_\alpha(h, f)
\end{align*}
Since in the last equation both factors of $\nabla_{\alpha, \alpha}(f\circ g)$ belong the image of $\mathbf{U'}$, we may reduce $\nabla_{\alpha, \alpha}$ by it's universality to obtain (\ref{naturality}).

Next we should prove
\begin{equation}\label{mfassoc}
\phi_\alpha(g\circ h, f)\circ(\id_{f{\downarrow_{\alpha, \alpha}}}\star \phi_\alpha(h, g)) = \phi_\alpha(h, f\circ g)\circ(\phi_\alpha(g, f)\star\id_{h{\downarrow_{\alpha, \alpha}}})
\end{equation}
Composing the left part  of (\ref{mfassoc}) with $\nabla_{\alpha, \alpha}(f\circ g\circ h)$ and modifying using (\ref{mfdef}) twice, we have
\begin{align*}
\nabla_{\alpha, \alpha}(f\circ g\circ h)\circ\phi_\alpha(g\circ h, f)\circ(\id_{f{\downarrow_{\alpha, \alpha}}}\star \phi_\alpha(h, g)) = \\\nabla_{\alpha, \alpha}(f)\star\nabla_{\alpha, \alpha}(g)\star\nabla_{\alpha, \alpha}(h)
\end{align*}
The same result ($\nabla_{\alpha, \alpha}(f)\star\nabla_{\alpha, \alpha}(g)\star\nabla_{\alpha, \alpha}(h)$) can be obtained from the right part of (\ref{mfassoc}) composed with $\nabla_{\alpha, \alpha}(f\circ g\circ h)$ using similar procedures. It means that
\begin{align*}
\nabla_{\alpha, \alpha}(f\circ g\circ h)\circ\phi_\alpha(g\circ h, f)\circ(\id_{f{\downarrow_{\alpha, \alpha}}}\star \phi_\alpha(h, g)) =\\ \nabla_{\alpha, \alpha}(f\circ g\circ h)\circ\phi_\alpha(h, f\circ g)\circ(\phi_\alpha(g, f)\star\id_{h{\downarrow_{\alpha, \alpha}}})
\end{align*}
But this factor $\nabla_{\alpha, \alpha}(f\circ g\circ h)$ can be reduced by it's universality and we obtain (\ref{mfassoc}).

Finally we need following equations:
\begin{align}\label{mfid}
&\phi_\alpha(\id_A, f)\circ(\id_{f{\downarrow_{\alpha, \alpha}}}\star\phi_{\alpha,0}) = \id_{f{\downarrow_{\alpha, \alpha}}}\\
&\phi_\alpha(f, \id_A)\circ(\phi_{\alpha,0}\star\id_{f{\downarrow_{\alpha, \alpha}}}) = \id_{f{\downarrow_{\alpha, \alpha}}}
\end{align}
Prove only last one, because proofs are similar. Composing the left part with $\nabla_{\alpha, \alpha}(f)$ and modifying it with (\ref{mfdef}) we obtain $(\nabla_{\alpha, \alpha}\circ\phi_{\alpha, 0})\star\nabla_{\alpha, \alpha}(f)$, which is equal to $\alpha\star\nabla_{\alpha, \alpha}(f)$ by (\ref{mf0def}). But by (\ref{nabladef}) it is equal to $\nabla_{\alpha, \alpha}(f)$. And again we may reduce $\nabla_{\alpha, \alpha}(f)$ to obtain the desirable result.
\item The proof of this part of proposition is very similar to proofs of previous parts, so we don't repeat that.$\square$
\end{enumerate}

\end{document}